\documentclass[%
 reprint,
superscriptaddress,
 amsmath,amssymb,
 aps,prl
]{revtex4-1}

\usepackage[utf8]{inputenc}
\usepackage[english]{babel}
\usepackage{graphicx}% Include figure files
\usepackage{dcolumn}% Align table columns on decimal point
\usepackage{bm}% bold math
\usepackage{braket}
\usepackage{color}
\newcommand{\im}{\operatorname{i}}

\DeclareMathOperator*{\Tr}{Tr}
\allowdisplaybreaks

\begin{document}

\preprint{APS/123-QED}

\title{Adsorption energies of benzene on close packed transition metal surfaces using the random phase approximation}% Force line breaks with \\
\author{Jos\'e A. Garrido Torres}
\affiliation{%
 EaStCHEM and School of Chemistry, University of St Andrews, KY16 9ST, St Andrews, UK
}%
\author{Benjamin Ramberger}
\affiliation{%
 Faculty of Physics and Center for Computational Materials
Sciences, University of Vienna, Sensengasse 8/12, 1090 Wien, AT
}%

\author{Herbert A. Fr\"uchtl}
\affiliation{%
 EaStCHEM and School of Chemistry, University of St Andrews, KY16 9ST, St Andrews, UK
}%
\author{Renald Schaub}
\affiliation{%
 EaStCHEM and School of Chemistry, University of St Andrews, KY16 9ST, St Andrews, UK
}%
\author{Georg Kresse}
\email{georg.kresse@univie.ac.at}
\affiliation{%
 Faculty of Physics and Center for Computational Materials
Sciences, University of Vienna, Sensengasse 8/12, 1090 Wien, AT
}%

\date{\today}% It is always \today, today,

\begin{abstract}
The adsorption energy of benzene on various metal substrates is predicted using the random phase approximation (RPA) for the correlation energy.
Agreement with available experimental data is systematically better than 10\% for both coinage and reactive metals. 
The results are also compared with more approximate methods, including vdW-density functional theory (DFT), 
as well as dispersion corrected DFT functionals. Although dispersion corrected 
DFT can yield accurate results, for instance, on coinage metals, the adsorption energies are clearly overestimated
on more reactive transition metals. Furthermore, coverage dependent adsorption energies are well
described by the RPA. This shows that for the description of aromatic molecules on metal surfaces
further improvements in density functionals are necessary, or more involved many body
methods such as the RPA are required.
\end{abstract}

\pacs{Valid PACS appear here}% PACS, the Physics and Astronomy
\maketitle

%\tableofcontents
%----------------------------------------------------------------------------------------------------------
%\section{\label{sec:intro}Introduction}
%----------------------------------------------------------------------------------------------------------

The accurate prediction of adsorption energies of molecules on metal surfaces is a challenging subject
in condensed matter physics, physical chemistry, as well as applied catalysis research.
It is now well understood that although semi-local functionals generally predict  trends between different metal surfaces
reasonably well,  absolute adsorption energies even for prototypical molecules can be inaccurate \cite{Hammer1995,Norskov2008,Norskov2009,Feibelman2001}. 
This is particularly true for the very difficult cases of 
aromatic molecules, since their adsorption involves a mixture of covalent bonding
and van der Waals (vdW) bonding. The former is overestimated by the most commonly used DFT functional,
the Perdew-Burke-Ernzerhof (PBE) functional \cite{Perdew1996}, whereas the later is not captured at all by gradient corrected
functionals. Sometimes these two errors compensate, but most often this is not the case.
Progress in the inclusion of vdW corrections has been remarkable in the last years. The most successful schemes 
are additive D3 dispersion corrections by Grimme and coworkers\cite{Grimme2010,Grimme2011},
the similar scheme of Tkatchenko and Scheffler (TS) \cite{Tkatchenko2009}, as well
as van der Waals density functionals (vdW-DF) that depend on the density at two positions in space. The later were originally introduced by
Dion, Lundqvist and coworkers \cite{Dion2004}, but since the accuracy of the first functional was not always satisfactory,
many ``improved'' functionals emerged soon after (e.g. Ref. \onlinecite{Murray2009,Klimes2012,berland2015van}).
 
The adsorption of aromatic molecules has been thoroughly investigated using different functionals on an extensive range of (111) transition metal surfaces, 
 including coinage and catalytic substrates \cite{Yamagishi2001,Morin2004,Toyoda2011,Liu,Yildirim2013b,Carrasco2014,Miller2015}. 
 The results including vdW corrections are overall quite promising, however, a careful inspection of the numbers shows that they are not entirely
 satisfactory. In particular, on catalytic substrates, the adsorption energies for aromatic molecules tend
 to be overestimated. It is a simple matter to understand this issue. In the approximate methods described above,
 either a polarizability is assigned to individual atoms yielding predominantly pair-wise interactions between
any two atoms, or a vdW like interaction between any two points in space is introduced using the jellium gas as reference.
 The atom based partitioning neglects that in metal substrates, local fluctuations by $d$ electrons
 are strongly screened by the electrons at the Fermi-surface (Drude term). On the other hand,
 the jellium electron gas is not necessarily an accurate reference able to describe the 
 localized $d$ electron fluctuations. Thus the interplay between local polarizability and metallic screening is not  captured 
 by either of the two approximations. Likewise, both methods are combined with approximate density
 functionals that sometimes fail to describe covalent bonding contributions accurately, as for instance
 amply reported for CO on metal surfaces \cite{Feibelman2001,Kohler2004,Stroppa2008}.

The only seamless approach capable of describing local $d$ electron fluctuations, free electron like excitations across the Fermi-level,
as well as covalent bonding, is the random phase approximation to the correlation energy \cite{Harl_PRL_RPA_2009,Lebegue2010,Schimka2010}.
In this approximation,  first a DFT calculation is performed using an approximate density functional (here the PBE functional ~\cite{Perdew1996}).
Then the exact exchange and RPA correlation energy are evaluated as
\cite{Nozier_RPA_1958,Langreth_RPA_1977,Miyake_RPA_2002,Fuchs_RPA_2005,Furche2008,Harl_PRL_RPA_2009,Harl2010}:
 \begin{equation}
\small
  E = E_{\rm EXX} + \underbrace{\frac{1}{2 \pi} \int_0^\infty {\rm d} \nu \Tr[ \ln(1-\chi(\im\nu)\operatorname{v})+\chi(\im \nu)\operatorname{v}]}_{ E_{\rm RPA}}, \label{equ:RPA_energy}
 \end{equation}
where $E_{\rm EXX}$ is the Hartree-Fock energy functional determined using PBE orbitals, $\chi(\im\nu)$  is the independent particle  polarizability calculated using PBE  orbitals and one electron energies, and $\operatorname{v}$
is the Coulomb kernel.  Here, we use the RPA to evaluate the adsorption energy of benzene on 
various substrates. We show that only RPA yields reference quality results in excellent agreement with experiment.

\begin{figure}
		\includegraphics[width=85mm]{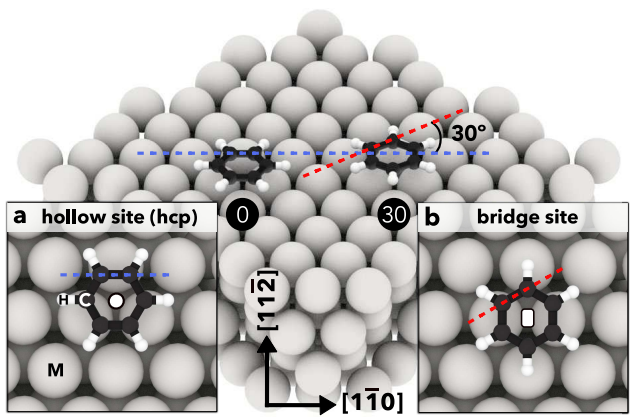}
	\caption{(color online) Representation of the two most common adsorption sites for benzene on transition metals. Magnifications from a top-view perspective for benzene adsorbed on the (a) 3-fold hcp and (b) 2-fold bridge sites. The blue and red dashed-lines are set to emphasise the different alignment of the C-C bonds for the 0 and 30 degrees configurations. Carbon, hydrogen and metal atoms are represented as black, white and grey spheres, respectively.}
	\label{fig:Bz-Ads-Sites}
\end{figure}

LEED and STM experiments as well as theory 
show that the benzene molecule adsorbs mainly at two sites, the 2-fold (bridge) or the 3-fold (hollow) site with  the C$_6$ ring parallel to the surface plane \cite{Neuber1995, Hove1986, Science1988, Schaff1996, Rockey2006, Witte1993,Carrasco2014, Liu2015,Yildirim2013b} (see
Fig. \ref{fig:Bz-Ads-Sites}). 
The preference for one or the other site is driven by a combination of different factors, such as the nature of the substrate \cite{Science1988}, the coverage \cite{Schaff1996} or the presence of co-adsorbates \cite{Witte1993}.
For coinage metals, the carbon-hydrogen planarity is not disturbed \cite{Syomin2001}, whereas the molecule is significantly deformed when strong chemisorption occurs \cite{Koel1984}. 

%-------------------------------------------------------------------------
%-------------------------------------------------------------------------

\begin{figure}
		\includegraphics[width=85mm]{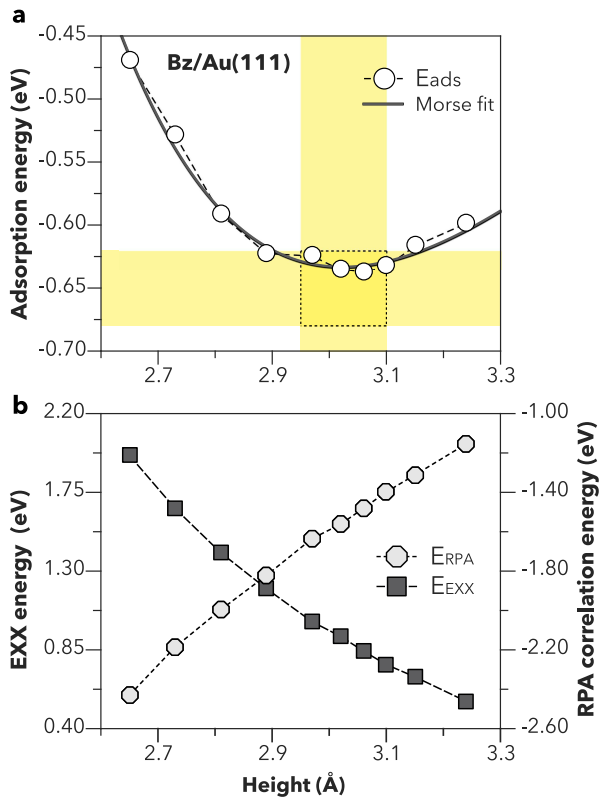}
	\caption{(color online)  (a) Plot of adsorption energy versus benzene height when adsorbed on Au(111). 
	The roughness is related to the use of a fairly coarse grid for the fast Fourier transformations in 
    the RPA total energy calculations. The yellow rectangles 
    mark the range of experimental values with uncertainties \cite{Liu, Liu2015}. The minimum of the energy curve lies within the experimental range (highlighted by the dashed rectangle) (b) Individual $E_{\rm EXX}$ and $E_{\rm RPA}$ contributions as a function of distance.
     Notice the large compensation between the two terms.}
	\label{fig:Au}
\end{figure}

Our strategy to calculate the adsorption energy is to fully relax the benzene molecule for each of the considered functionals for both adsorption registries (hcp, bridge). 
Although, we have recently presented a method to calculate forces between the atoms in the
RPA~\cite{ramberger2016analytic}, for efficiency reasons we have decided to use PBE  geometries for the
RPA calculations in the present work. The calculations were performed using the cubic scaling RPA of
Kaltak and coworkers \cite{Kaltak2014}.
For 
Cu, Ag and Au, we  found that a slightly more refined procedure was necessary to obtain accurate surface energies.
For these coinage metals, we varied the distance between the center of mass of the molecule and the substrate, relaxed all other degrees of freedom using
PBE, and then calculated the RPA energies for all considered molecule-surface distances. This procedure
was required, since none of the functionals yielded a satisfactory surface-molecule distance for benzene on these
metals. The optimized substrate-molecule distance for RPA was $\sim$~3.0~\AA, 2.95~\AA\ and 3.02~\AA \
for Cu, Ag, and Au, respectively. As an example, results for the energy-versus distance curve for Au 
are shown in Fig. \ref{fig:Au}.

%-------------------------------------------------------------------------
%-------------------------------------------------------------------------

We start with a brief discussion of the DFT results. The coinage metals are characterized by
a completely filled $d$ shell, and adsorption is essentially dominated by vdW contributions.
Since the PBE functional
does not account for vdW interactions, the predicted adsorption energies on the coinage metals are smaller than 0.1~eV for PBE. 
Furthermore,  the predicted surface-molecule distances are typically about 3.8~\AA, which are clearly much larger than the values
obtained from the RPA (compare also Fig. \ref{fig:Au}). On the open shell $d$ metals (Ni, Pd, Pt and Rh), adsorption energies of about 1.0~eV are predicted, 
which are about 0.8~eV smaller than the experimental values. 
The atom-based vdW correction schemes (PBE-D3\cite{Grimme2010} and PBE-TS \cite{Tkatchenko2009})
behave remarkably similar, with corrections of 0.7~eV for coinage metals and  1.2~eV for the considered reactive metals. The Ni substrate is a clear outliner, with corrections
amounting to 2~eV. This  indicates that there are issues with the parametrization of vdW corrections for magnetic
Ni, and possibly magnetic transition metals in general.  
The optB86b-vdW functional also yields reasonable results, with corrections of 0.6~eV for coinage metals, 
and 1.25~eV for the reactive metals. The values are seemingly somewhat more consistent than for the 
PBE-D3 and PBE-TS cases. In particular, for benzene adsorption on Ni, the corrections are now in line with the other
transition metals.
Results for the vdW-DF \cite{Dion2004} and vdW-DF2 \cite{Murray2009} functionals are 
disappointing. The adsorption energies behave in a very non-systematic manner. Even, if we exclude
the magnetic Ni (neither of the two functionals are suitable for magnetic substrates\cite{thonhauser2015spin}),
meaningful improvements compared to PBE can only be observed for the coinage metals. 

%-------------------------------------------------------------------------
\begin{table}[h]
		\caption {Calculated adsorption energies (in eV) for benzene adsorbed on different (111) metal substrates at medium coverage. Experimental values from the literature are included for comparison. Calculations were performed for a  (2$\sqrt3$ $\mathsf{x}$ 4) unit cell with 4 layers. The topmost 2 layers were relaxed. 3$\mathsf{x}$3$\mathsf{x}$1 \textit{k}-points were used for the RPA, and corrected for k-point errors using the PBE functional
		(see supplementary material for details). 
		} \label{tab:Bz-RPA-Metals} 
 	\begin{ruledtabular}
	\begin{tabular}{l ccccccc} 
                           & Cu & Ag & Au & Ni & Pd & Pt & Rh  \\
                           \hline
	PBE    & 0.09	& 0.05	& 0.05	& 1.04	& 1.16	& 0.82	& 1.46 \\ 
	PBE-D3    & 0.98  & 0.78	& 0.83	&	3.09&	2.34 &	2.09 &	2.58 \\
	PBE-TS  & 0.87 & 0.77	 & 0.80 & 2.85 &	2.10&	1.99&	2.72 \\
	vdW-DF & 0.53	& 0.46	 & 0.53	& 0.56 & 0.83 & 0.37 & 1.16 \\
	vdW-DF2 &	0.75&0.43	&0.43&	$-$0.07&	 0.63&	1.33 &	0.63 \\
	optB86b-vdW   &	0.69	 & 0.69 &	0.76 & 2.25	& 2.41	& 2.12	& 2.68 \\
	RPA     &0.66&	 0.63&	0.64 &	1.46&	1.72&	1.74&	2.08 \\
	       \hline
	Exp. &	0.69$^{a}$& 0.68$^{a}$ &	0.65$^{a}$&	-&	1.74$^b$&	1.67$^c$&	- \\
	$\Delta$ Exp. &	$\pm$0.04 & $\pm$0.05 &	$\pm$0.03 & -	 &$\pm$0.30&	$\pm$0.17&	- 
	\end{tabular}
	\end{ruledtabular}
	\begin{flushleft}
   $^a$ Ref. \cite{Liu2015}; recent interpretation of TPD spectra based on the Polanyi-Wigner equation \cite{King1975}. \\
   $^b$ Ref. \cite{Yildirim2013b}; TPD experiments using the Redhead analysis \cite{Redhead1962}. \\
   $^c$ Ref. \cite{Ihm2004}; microcalorimetry measurements for benzene at  $\theta$=0.80 ML, corresponding to the 2$\sqrt3$ $\mathsf{x}$ 4 unit cell.
     \end{flushleft}
\end{table}
%-------------------------------------------------------------------------

The best agreement with experiment is obtained for the RPA, in which every single predicted value falls within the experimental error bars.
 Having the RPA at hand, we are now in a much better position to validate the other functionals, since the error bars
in the experimental values are simply too large to make a conclusive statement
on the accuracy of the vdW correction schemes relying on experiments only. RPA suggests that PBE-D3, PBE-TS and optB86b-vdW
overestimate the vdW corrections slightly for coinage metals and substantially for reactive 
metals. By this statement we mean that the correction (the difference between PBE and PBE-D3)
is up to twice as large as the difference between RPA and PBE. But this error is also non-systematic:
for Cu, Ag, Au, and Pt, the error seems to be only about 25\%, however it increases to almost 100\% 
for Pd and Rh. It is not unlikely that a substantial part of the error is already present in the
parent DFT functional. But it is equally well possible that the error is related to the neglect
of the interplay between the local excitations and jellium like screening in metals. If one neglects
the metallic screening, the interaction between the substrate atoms and the molecule is expected
to be overestimated \cite{rehrkohn1975van,Zaremba1976vdW}. 
Inclusion of this effect  slightly improves agreement with experiment,
but still yields too large adsorption energies (Pd 2.14~eV, Rh 2.52~eV)~\cite{Liu}.
Many body dispersion corrections \cite{tkatchenko2012MBD} provide a more systematic approach to resolve
this issue. However, adoption to metals is not straightforward, and even then
the issue of the accuracy of the underlying semi-local DFT functional remains to be addressed.

%-------------------------------------------------------------------------
\begin{figure}
		\includegraphics[width=85mm]{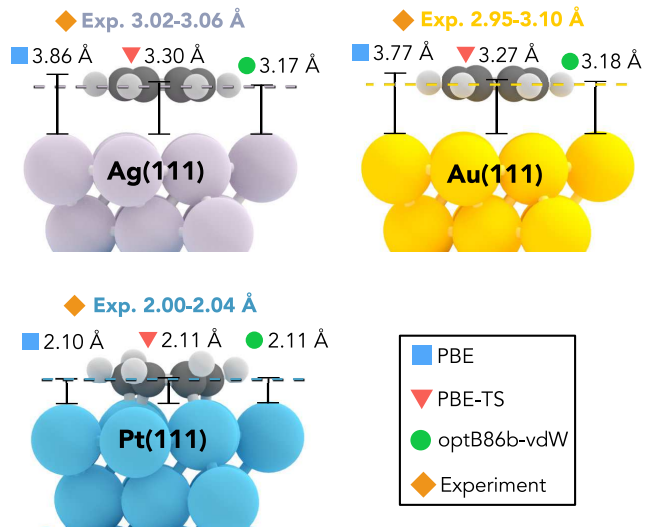}
	\caption{(color online) 
	Averaged carbon-metal distances for benzene on  Ag, Au and Pt using the PBE, PBE-TS
and optB86b-vdW functionals.  Experimental heights are included for comparison Ag \cite{Liu2015}, Au\cite{Liu, Liu2015} and Pt \cite{Wander1991, Ihm2004}. \label{fig:geometry}}
\end{figure}
%-------------------------------------------------------------------------

We now  comment briefly on the geometries. To this end we show in Fig. \ref{fig:geometry} 
results for Ag, Au, and Pt, for which experimental 
geometries are accurately known.  It is clear that PBE predicts a much too large benzene-substrate distance on Ag and Au, but
even the optB86b-vdW functional yields distances that are somewhat larger than experimentally measured. The value
of  2.95~\AA\ (3.02~\AA) obtained with the RPA for Ag (Au) is very close to the experimental range of values. 
As an example for strong chemisorption, we also show the results for Pt, where we observe that
the different functionals yield very similar benzene substrate distances. We also compared 
RPA adsorption energies determined using different geometries, for instance the optB86b-vdW geometries,
and found little variations for the energies between the geometries. 
Furthermore, an optimization of the surface-benzene distance for Pt, in the same manner as for the coinage metals, 
changed the results very little. 
This is an {\em a posteriori} justification for our choice to
use PBE geometries for the RPA calculations for reactive substrates. 
In summary, for coinage metals the geometries change substantially if vdW contributions
are taken into account and only the RPA yields entirely satisfactory results, whereas we see little 
variations between functionals for reactive metals.

%-------------------------------------------------------------------------
\begin{figure}
		\includegraphics[width=85mm]{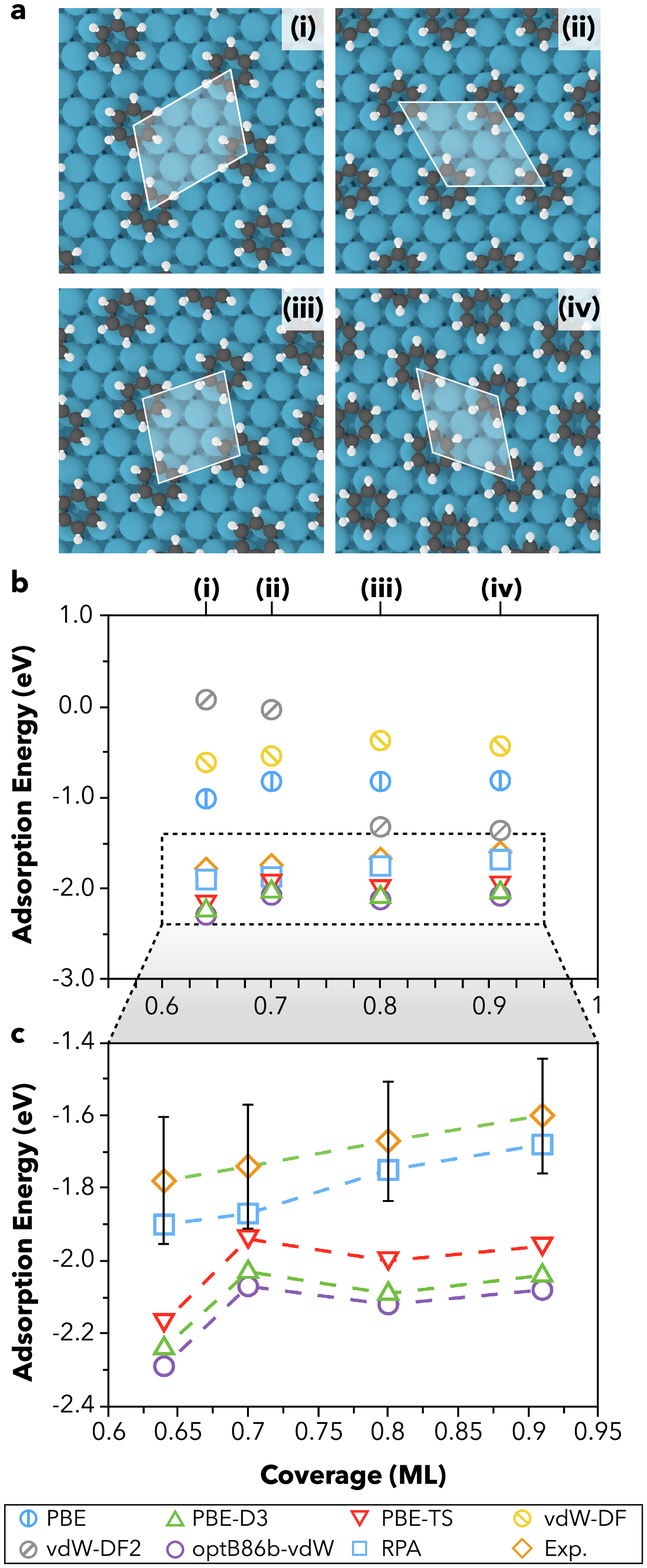}
	\caption{(color online)  
	Adsorption energy as function of the coverage for benzene on Pt(111). (a) Considered geometries for benzene at coverages $\theta_{\text{i}}$=0.64, $\theta_{\text{ii}}$=0.71, $\theta_{\text{iii}}$=0.80 and $\theta_{\text{iv}}$=0.91 monolayers (ML), defined by the corresponding surface unit cells: (i)=$\scriptsize \begin{pmatrix} 4&2&|&$-$1&3 \end{pmatrix}$, (ii)=$\scriptsize\begin{pmatrix} 3&0&|&0&3 \end{pmatrix}$, (iii)=$\scriptsize\begin{pmatrix} 3&1&|&1&3 \end{pmatrix}$ and  (iv)=$\scriptsize\begin{pmatrix} 2&$-$1&|&1&3 \end{pmatrix}$. (b) Adsorption energy (in eV) for benzene on Pt(111) for those coverages (in ML). (c) Magnification of the previous plot in the area near the experimental energy range (dotted lines are added to underline the trends). Integral of the differential heat of adsorption and error bars, corresponding to the spread of the experimental results, are included from Ref. \cite{Ihm2004}. \label{fig:coverage}
	For the coverage, we have adopted the convention of Ihm and coworkers\cite{Ihm2004}. 	}
\end{figure}
%-------------------------------------------------------------------------

The final issue we would like to address in this letter is whether the RPA also improves
the description of the coverage dependence of the adsorption energy. 
Microcalorimetry experiments have been performed by the group of Campbell \cite{Ihm2004} to investigate this dependence 
for benzene on Pt(111). Panel (b) in Fig. \ref{fig:coverage} shows the results for all functionals, 
including PBE, vdW-DF and vdW-DF2. As before, results for these three functionals are
not satisfactory on an absolute scale.  We therefore disregard them in further discussion
and concentrate on the results for the more accurate functionals in panel (c).
The RPA shows a steady decrease of the average adsorption energy (corresponding
to the integral of the differential heat of adsorption). This is expected, because of
repulsion between the molecules  as well
as a progressive passivation of the substrate atoms \cite{Norskov2008}.
An issue that all density functionals share is that structure (II) is unstable, since it lies above the connecting line
between phase (I) and (III). Interestingly, the RPA predicts this phase to be stable.
 It is clear that RPA matches the experimental coverage dependence very well, falling
always within the experimental error bars. But we note that the error
in the vdW corrected functionals is mostly in the absolute value of the adsorption energy
and less so in the repulsive interaction between the molecules. 

In summary, we have calculated the adsorption energy of benzene on coinage metals
as well as prototypical catalytic substrates.  Agreement with the available experimental
data for adsorption energies is excellent, both at the lowest considered coverage and at a  range of coverages for Pt. 
Furthermore, for coinage metals the RPA predicts benzene-substrate distances of about 3.0~\AA\ in perfect agreement with experiment, 
significantly smaller than for semi-local functionals and somewhat smaller than for most vdW corrected functionals
(3.2~\AA). As computations are very affordable (on 192 cores a calculation required typically two to three hours),
reference results for adsorption on metal surfaces can now be easily and reliably obtained
even for complex molecules with mixed covalent and van der Waals bonding. This is an important
step towards an accurate first principles modelling of substrate-adsorbate interactions.
The present study establishes that a similar systematic improvement as for CO on metal surfaces\cite{Schimka2010} 
carries over to the much more challenging aromatic molecules.

{\em Acknowledgment:} We acknowledge financial support from the Scottish Funding Council (through EaStCHEM and SRD-Grant HR07003) and from EPSRC (PhD studentship for JAGT, EP/M506631/1).  Funding by the Austrian Science Fund (FWF): F41 (SFB ViCoM) is grateful
acknowledged.  Computations were predominantly performed on the Vienna Scientific Cluster VSC3.

\end{document}